\documentclass{iopart}
\usepackage{epsfig}
\usepackage{subfigure}
\def\bi{\bibitem}

\begin{document}
 
\article[Early dynamics of transversally thermalized matter]
  {Quark Matter 2008, Jaipur, India, February 2008}
  {Early dynamics of transversally thermalized matter\footnote{Research supported by the Polish Ministry of Science and Higher Education grants: 1 P03B 045 29 (2005-2008), N202~153~32/4247 (2007-2009) and 
  N202~034~32/0918 (2007-20010). }
}  
 
\author{A. Bialas$^{1,2}$, M. Chojnacki$^2$,  W. Florkowski$^{2,3}$ }

\address{$^1$ M.~Smoluchowski Institute of Physics, Jagellonian University, 30-059 Krak\'ow, Poland}
\address{$^2$ The H.~Niewodnicza\'nski Institute of Nuclear Physics, 
Polish Academy of Sciences, 31-342 Krak\'ow, Poland} 
\address{$^3$ Institute of Physics, Jan~Kochanowski University, 25-406 Kielce, Poland}

\ead{Mikolaj.Chojnacki@ifj.edu.pl}
   
\begin{abstract}
We argue that the idea that the parton system created in relativistic heavy-ion collisions  is formed in a state with transverse momenta close to thermodynamic equilibrium and its subsequent dynamics at
early times is dominated by pure transverse hydrodynamics of the perfect fluid is compatible with the data collected at RHIC. This scenario of early parton dynamics may help to solve the problem of early equilibration. 
\end{abstract}

\bigskip

It is now commonly accepted that the evolution of matter created in heavy-ion collisions at RHIC energies is most successfully described by hydrodynamics of the perfect fluid \cite{Shuryak:2004cy,Kolb:2003dz,Huovinen:2003fa}. In particular, the hydrodynamical approach reproduces reasonably well the particle transverse-momentum spectra and the elliptic flow coefficient $v_2$. Very recently, it has been also shown that the hydrodynamic model can describe consistently the HBT radii \cite{Broniowski:2008vp}. 

On the other hand, the hydrodynamic picture is challenged by a serious problem. To reproduce correctly the data, the hydrodynamic evolution should start at the time well below 1 fm after the collision takes
place. Such fast equilibration time is not easy to achieve with elastic perturbative
cross-sections, hence the success of the hydrodynamic approach inevitably leads to the puzzle of early thermalization.  This puzzle was widely discussed and several exotic mechanisms were proposed for its solution \cite{Mrowczynski:2005ki,Strickland:2007fm}, however no one was yet accepted as fully satisfactory.

In the present paper we analyze the possibility that, at the early stages of high-energy collisions, the hydrodynamic evolution applies only to transverse degrees of freedom of the partonic system. The longitudinal dynamics is essentially described by partonic free-streaming (we consider here the midrapidity region, $y \approx 0$). 

\begin{figure*}[t]
\begin{center}
\includegraphics[angle=0,width=0.80\textwidth]{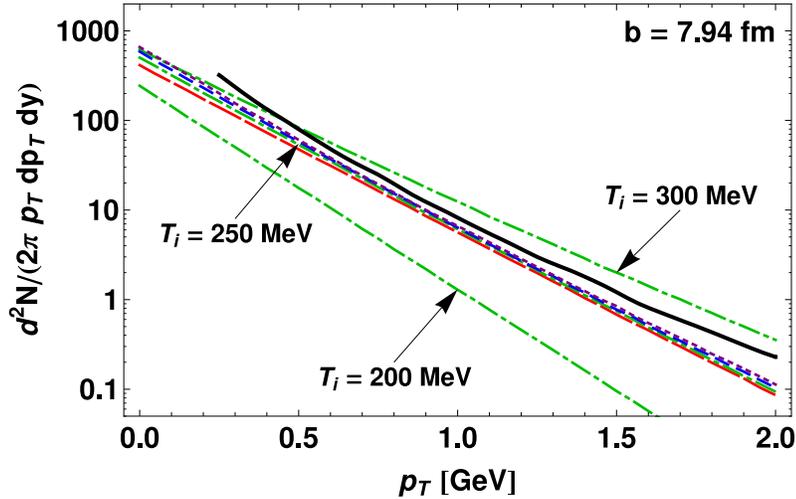}
\end{center}
\caption{\small {Transverse-momentum spectra of $\pi^+$ measured by  PHENIX  
 in the centrality class 30-40\% (solid line) \cite{RHIC-spectra} and the model spectra of gluons for various choices of $T_i$ and $T_f$. The lowest dash dotted curve corresponds to $T_i$ = 200 MeV and $T_f$ = 180 MeV. The highest dash dotted curve was obtained for $T_i$ = 300 MeV and $T_f$ = 180 MeV. The four almost parallel lines represent our results for $T_i$ = 250 MeV and for four final values of the temperature: $T_f$ = 200 (long dashed line), 180 (dash dotted line), 160 (dashed line) and 140 (dotted line) MeV. Note that the original $\pi^+$ experimental spectra collected at $\sqrt{s_{NN}}$ = 200 GeV have been multiplied by a factor of 3 to account for the total hadron multiplicity. }}
\label{fig:spectra}
\end{figure*}

The main reason for our study is the observation that the early equilibration, if at all possible, is particularly difficult to achieve in longitudinal direction. On the other hand,  the equilibration of the partonic transverse-momentum spectrum may be obtained much easier. In fact, it is well known that the transverse momentum spectra observed in nucleon-nucleon collisions are well described by the Boltzmann
distribution \cite{bec1,bec2}.

In our approach the partonic system may be treated as a superposition of independent transverse clusters. The clusters are formed by the particles having the same value of the rapidity. The space-time evolution of the system is described by the hydrodynamic equations introduced in Ref. \cite{Bialas:2007gn}. They are based on the energy and momentum conservation laws
\begin{equation}
\partial_\mu T^{\mu \nu} = 0,
\label{enmomcons}
\end{equation}
where the energy-momentum tensor is defined by the formula 
\begin{eqnarray}
T^{\mu \nu} &=& \frac{n_0 \nu_g  T^3}{2\pi \tau} \left( 3 U^\mu U^\nu -g^{\mu \nu} - V^\mu V^\nu \right).
\label{Tmunu}  
\end{eqnarray} 
In Eq. (\ref{Tmunu}) the parameter $n_0$ is the density of clusters in rapidity space, $\nu_g$ is the degeneracy factor (we assume that our system is dominated by gluons, hence we use $\nu_g$ = 16), $T$ is the temperature, and $\tau=\sqrt{t^2-z^2}$. The four vectors $U^\mu$ and $V^\mu$ are defined by the equations:
\begin{equation}
U^{\mu} = ( u_0 \cosh\eta,u_x,u_y, u_0 \sinh\eta),
\label{U}
\end{equation}
\begin{equation}
V^{\mu} = (\sinh\eta,0,0,\cosh\eta),
\label{V}
\end{equation}
where $\eta$ is the spacetime rapidity, and $u^0, u_x, u_y$ are the components of the four-velocity of the fluid in the rest-frame of a cluster, where $y=\eta=0$,
\begin{equation}
u^\mu = \left(u^0, {\vec u}_\perp, 0 \right) = \left(u^0, u_x, u_y, 0 \right).
\label{smallu}
\end{equation}
One of the interesting features of our approach defined by Eqs. (\ref{enmomcons}) - (\ref{smallu}) is that it naturally leads to the entropy conservation, which takes the form
\begin{equation}
\partial_\mu S^\mu = 0, 
\label{entcon}
\end{equation}
where
\begin{equation}
S^\mu =  \frac{3 n_0 \, \nu_g T^2}{2\pi \tau} U^\mu.
\label{smu}
\end{equation}
Eq. (\ref{entcon}) follows directly from the projection of Eq. (\ref{enmomcons}) with the energy-momentum tensor (\ref{Tmunu}) on the four-velocity $U_\nu$. More information concerning the  formal aspects of transverse hydrodynamics may be found in Refs. \cite{Chojnacki:2007fi,Ryblewski:2008fx}. Here we only note that our approach differs from the similar model by Heinz and Wong \cite{hw}, where the entropy is not conserved and the system is treated as dissipative.

\begin{figure*}[t]
\begin{center}
\includegraphics[angle=0,width=0.80\textwidth]{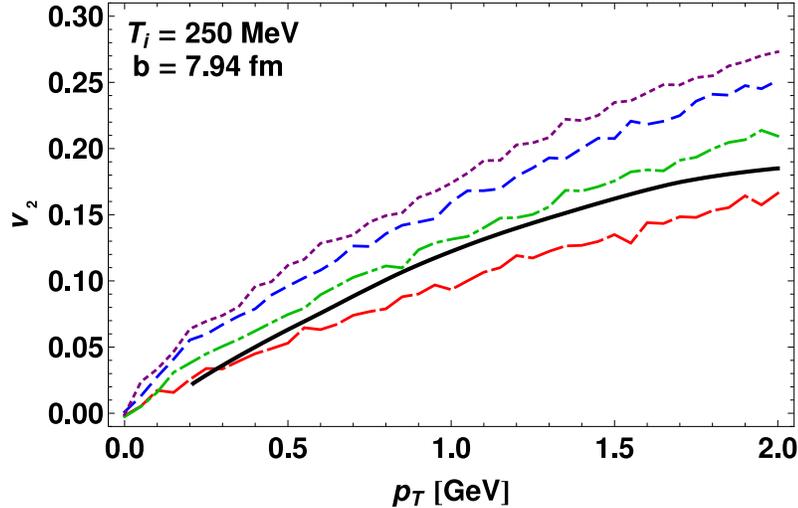}
\end{center}
\caption{\small {The elliptic flow coefficient $v_2$ as the function of the transverse momentum. The PHENIX experimental results for pions and kaons in the centrality class 20-40\% and for the collision energy $\sqrt{s_{NN}}$ = 200 GeV \cite{RHIC-v2} are compared to the model calculations with $T_i$ = 250 MeV and for four final values of the temperature \mbox{$T_f = 200,$} 180, 160 and 140 MeV. The curves are denoted in the same way as in Fig. 1. The best agreement is obtained for $T_f \approx$ 180 MeV.}}
\label{fig:ptv2}
\end{figure*}

To solve numerically the hydrodynamic equations (\ref{enmomcons})  we have to specify the initial conditions, appropriate for the physical situation encountered in gold on gold collisions at RHIC energy.
In our approach we assume that the transverse profile of the initial energy density is given by the density of wounded nucleons \cite{wn} (see also \cite{abwc,bb}). This density is determined, for a given centrality, from the Glauber model. In the theoretical calculations we consider the centrality class $20-40$\%, corresponding to the impact parameter $b\approx$ 7.9 fm. The initial temperature $T_i$ at the origin, ${\vec x}_\perp = 0$, is taken as a free parameter.

The parton spectra and the elliptic flow coefficient $v_2$ are evaluated using the Cooper-Frye prescription \cite{fc}
\begin{equation}
\frac{dN}{d^2p_\perp dy} =\frac {n_0 \, \nu_g}{(2\pi)^2} \int d\Sigma_\mu(x)
p^\mu \, F(x,p),
\end{equation}
where $\Sigma$ is the hypersurface where the purely transverse evolution comes to the end, and $F(x,p)$ is the underlying phase-space distribution function, see Ref. \cite{Bialas:2007gn}. In our calculations $\Sigma$ is determined by the condition of constant temperature $T=T_f$. The details of our calculation of the parton transverse-momenta and the elliptic flow coefficient $v_2$ are given in Ref. \cite{Chojnacki:2007fi}.

Our results obtained for several choices of the parameters $T_i$ and $T_f$ are shown in Figs. 1 and 2.
We find that with $T_i \sim$ 250 MeV the slope of the transverse momentum spectrum may be easily adjusted to the one measured for the pion spectra \cite{RHIC-spectra}.  On the other hand, the experimental values for $v_2$ \cite{RHIC-v2} select the final-temperature value $T_f$ = 180 MeV.
This value of the final temperature is reached by a system in a relatively short evolution time of about 4 fm. The short evolution time implies that the transverse size of the system is relatively small, smaller than that obtained from the HBT measurements, thus consistently leaving space for further,
three-dimensional expansion of the system. 

We thus conclude that the concept of the initial transverse equilibration and hydrodynamic evolution of the partonic system is compatible with the data and the proposed model represents a possible solution the the problem of early equilibration.

Very recently, the successful description of the RHIC data has been achieved in a model which assumes initial free-streaming in three space directions followed by a sudden transition to standard hydrodynamics \cite{Broniowski:2008vp}. Similarly to our arguments, that model indicates that the successful description of the data may be achieved in the approach without complete thermalization in the early stage. Thus, further investigations of early non-equilibrium dynamics (for example, see Ref. \cite{Bozek:2007di}) are certainly necessary and interesting -- they bring us closer to the solution of the early thermalization puzzle.

\end{document}